\documentclass[sort&compress,final,5p,times,twocolumn]{elsarticle}
\usepackage{amssymb}
\usepackage{lipsum}
\usepackage{amsmath}

\journal{Physics Letters B}

\usepackage{mathpazo}
\usepackage{booktabs}
\usepackage{comment}
\usepackage{rotating}
\usepackage{graphicx,bm,color,xcolor}
\usepackage[hidelinks]{hyperref}
\hypersetup{
    colorlinks,
    linkcolor={red!50!black},
    citecolor={blue!50!black},
    urlcolor={blue!80!black}
}
\usepackage{cancel}

\newcommand{\ce}[1]{Eq.~(\ref{#1})}

\newcommand{\ct}[1]{{Table~\ref{#1}}}

\usepackage[caption=false]{subfig}

\usepackage{calligra}
\usepackage[T1]{fontenc}

\usepackage{blindtext}

\newcommand{\MG}{{\tt MG5aMC}\xspace}

\newcommand{\sqrtsnn}{\ensuremath{\sqrt{s_{_{NN}}}}}
\newcommand{\pT}{\ensuremath{P_T}\xspace}
\newcommand{\jpsi}{\ensuremath{J/\psi}\xspace}

\usepackage[normalem]{ulem}

\DeclareMathAlphabet{\pazocal}{OMS}{zplm}{m}{n}

\DeclareFontFamily{OT1}{pzc}{}
\DeclareFontShape{OT1}{pzc}{m}{it}{<-> s * [1.10] pzcmi7t}{}
\DeclareMathAlphabet{\mathpzc}{OT1}{pzc}{m}{it}

\usepackage{scalerel}
\usepackage{tikz}
\usetikzlibrary{svg.path}
\usepackage{academicons}
\definecolor{orcidlogocol}{HTML}{A6CE39}
\usepackage{orcidlink}

\tikzset{
  orcidlogo/.pic={
    \fill[orcidlogocol] svg{M256,128c0,70.7-57.3,128-128,128C57.3,256,0,198.7,0,128C0,57.3,57.3,0,128,0C198.7,0,256,57.3,256,128z};
    \fill[white] svg{M86.3,186.2H70.9V79.1h15.4v48.4V186.2z}
                 svg{M108.9,79.1h41.6c39.6,0,57,28.3,57,53.6c0,27.5-21.5,53.6-56.8,53.6h-41.8V79.1z M124.3,172.4h24.5c34.9,0,42.9-26.5,42.9-39.7c0-21.5-13.7-39.7-43.7-39.7h-23.7V172.4z}
                 svg{M88.7,56.8c0,5.5-4.5,10.1-10.1,10.1c-5.6,0-10.1-4.6-10.1-10.1c0-5.6,4.5-10.1,10.1-10.1C84.2,46.7,88.7,51.3,88.7,56.8z};
  }
}

\newcommand\orcidicon[1]{\href{https://orcid.org/#1}{\mbox{\scalerel*{
\begin{tikzpicture}[yscale=-1,transform shape]
\pic{orcidlogo};
\end{tikzpicture}
}{|}}}}

\begin{document}

\begin{frontmatter}

\title{The impact of future $D$- and $B$-meson measurements with the SMOG2 program at LHCb on the determination of nuclear parton distribution functions}

\author[addr1,addr2]{Carlo Flore \orcidlink{0000-0002-1071-204X}}
\ead{carlo.flore@unica.it}

\author[addr3]{Cynthia Hadjidakis \orcidlink{0000-0002-9336-5169}}
\ead{cynthia.hadjidakis@ijclab.in2p3.fr}

\author[addr4]{Daniel Kiko\l{}a \orcidlink{0000-0001-6896-6475}}
\ead{daniel.kikola@pw.edu.pl}

\author[addr5]{Aleksander Kusina \orcidlink{0000-0002-4090-0084}}
\ead{aleksander.kusina@ifj.edu.pl}

\author[addr4]{Anton Safronov\corref{cor1} \orcidlink{0000-0002-0586-0830}}
\ead{anton.safronov.dokt@pw.edu.pl}
\cortext[cor1]{Corresponding author}

\affiliation[addr1]{organization={Dipartimento di Fisica, Università di Cagliari},
             addressline={Cittadella Universitaria},
             city={Cagliari},
             postcode={I-09042},
             state={Monserrato (CA)},
             country={Italy}}

\affiliation[addr2]{organization={INFN, Sezione di Cagliari},
             addressline={Cittadella Universitaria},
             city={Cagliari},
             postcode={I-09042},
             state={Monserrato (CA)},
             country={Italy}}

\affiliation[addr3]{organization={Universit\'e Paris-Saclay, CNRS/IN2P3, IJCLab},
(            city={Orsay},
            postcode={91405}, 
            country={France}}

\affiliation[addr4]{organization={Faculty of Physics, Warsaw University of Technology},
            addressline={plac Politechniki 1}, 
            city={Warszawa},
            postcode={00-661}, 
            country={Poland}}

\affiliation[addr5]{organization={Institute of Nuclear Physics, Polish Academy of Sciences},
            addressline={ul. Radzikowskiego 152}, 
            city={Cracow},
            postcode={31-342}, 
            country={Poland}} 

\begin{abstract}
We perform an analysis of the potential impact of future $D$- and $B$-meson measurement within the SMOG2 fixed-target program at the LHCb experiment on nuclear parton distribution functions. Following~\cite{Bursche:2018orf}, we assume that SMOG2 will collect data for five nuclear targets: He, Ne, Ar, Kr, Xe and hydrogen which will provide a baseline for constructing nuclear ratios. The analysis is performed by using the PDF reweighting method.
We demonstrate that such a measurement will allow to considerably reduce the current uncertainties of the nuclear gluon distribution and, to some extent, even the uncertainties of the light sea-quark distributions. Furthermore, it will provide the first possibility to systematically study the current assumptions on the $A$-dependence of the nuclear parton distributions.
\end{abstract}


\end{frontmatter}

\section{\label{sec:introduction}Introduction}
One of the crucial aspects of high-energy nuclear physics is studying the structure of matter, in particular, in terms of its smallest components, quarks and gluons, collectively referred to as partons. 
Within the collinear factorisation approach~\cite{Bodwin:1984hc,Collins:1985ue}, the partonic structure of a hadron is described by parton distribution functions (PDFs), which represent the probability of finding partons with a given fraction $x$ of the proton longitudinal momentum. PDFs encapsulate the non-perturbative information about hadron content, and are determined through fits to experimental data~\cite{Ethier:2020way,Gao:2017yyd,Kovarik:2019xvh,Klasen:2023uqj}.

Furthermore, the partonic structure of nucleons bound in a nucleus differs from that of a free proton. This phenomenon was first identified as the EMC effect~\cite{EuropeanMuon:1981gyc}. 
Subsequent studies have shown that this modification varies with $x$ and we can now recognise four characteristic regions of nuclear modifications~\cite{EuropeanMuon:1988lbf,EuropeanMuon:1988tpw,EuropeanMuon:1983wih,Goodman:1981hc,Bodek:1983ec,BCDMS:1985dor,Klein:2019qfb}. These are identified as:
(i)~the shadowing region at $x\lesssim0.1$ (characterised by a depletion of nuclear cross-sections compared to the free-nucleon ones),
(ii)~the antishadowing region at $0.25\lesssim x \lesssim0.1$ (characterised by a corresponding enhancement),
(iii)~the EMC region $0.75 \lesssim x \lesssim 0.25$ (characterised by depletion),
and (iv)~the Fermi-motion region $x \gtrsim 0.75$ (characterised by enhancement).
The strength of these effects also changes with the atomic mass number of the nucleus.

Factorisation framework directly links these modifications observed on the level of cross sections with the behaviour of nuclear PDFs (nPDFs).
Therefore, a precise determination of nPDFs is essential for using perturbative quantum chromodynamics (pQCD) to interpret experimental data from high-energy nuclear reactions.  For example, this information is crucial for using charm and bottom mesons to study the hot and dense partonic matter created in ultrarelativistic heavy-ion collisions at the Relativistic Heavy Ion Collider at Brookhaven National Laboratory~\cite{STAR:2018zdy} or at the Large Hadron Collider (LHC) at CERN (for example~\cite{ALICE:2022wpn}), as well as for computing cross sections in astrophysics, e.g.~\cite{Bhattacharya:2016jce}. Despite significant progress over the past decades, our understanding of nPDFs, especially the gluon ones, remains incomplete. For instance, there is a limited knowledge about how nuclear effects vary with the number of nucleons in a nucleus.

Charm and bottom quarks in high-energy reactions are predominantly produced via gluon fusion. Therefore, the yields of charm and bottom hadrons can be used to determine gluon nPDFs~\cite{Kusina:2017gkz,Kusina:2020dki}. The LHCb experiment facilitates an effective measurement of $D$ and $B$ mesons in proton-nucleus collisions across various ion spe\-cies making it a perfect place for robust study on nuclear gluon PDFs.

In this paper, we review the prospects for investigating nuclear parton distribution functions using $p$A pseudo-data at $\sqrt{s_{NN}} = 113$ GeV from the SMOG2 fixed-target program at the LHCb experiment~\cite{BoenteGarcia:2024kba}.  The pseudo-data used for the analysis are generated with {\tt MadGraph5\_aMC@NLO}~\cite{Flore:2025ync}. 
The paper is organised as follows: in Section~\ref{sec:FT-LHCb} we begin with a brief overview of the main features of fixed-target collisions using LHC beams. Next, we describe the experimental setup for these collisions at the LHCb experiment, including the ion species that can be used and the expected parameters of this experimental program. Section~\ref{sec:formalism} introduces the formalism employed to evaluate the impact of the anticipated LHCb data on nPDFs. Finally, we present our results in Section~\ref{sec:results} and summarise our findings in Section~\ref{sec:conclusions}.

\section{Fixed-target program at the LHCb experiment\label{sec:FT-LHCb}}
\subsection{Overview of the fixed-target program at the LHC}
\label{subsec:FT-LHC}
An experiment utilising the LHC beams and a fixed target offers numerous, exciting and timely  opportunities~\cite{Hadjidakis:2018ifr,Brodsky:2012vg}. The high energy and intensity of the proton and ion beams at the LHC enable compelling studies of the partonic structure of the nucleus using charm- and bottom-quarks production, as well as measurements highly relevant to astrophysics. For the research considered in this paper, the key advantage is the ability to easily change the ion species of the target, and therefore to vary the mass number of the target nucleus over a wide range, from a single proton to heavy elements. Additionally, these collisions are expected to produce large yields of charmed mesons, ensuring a good statistical precision for studies involving charm quarks.
Furthermore, installing a polarised target would create unprecedented opportunities for studying the spin structure of a nucleon, which is a primary objective for the LHCspin initiative~\cite{Santimaria:2021zay,Pappalardo:2024jpe}.

The TeV-range energy of the LHC beams imposes specific requirements on the experimental setup. The LHC can provide a proton beam with energy up to 7 TeV, resulting in a center-of-mass energy for $p$A collisions of  $\sqrt{s_{NN}} = 115$ GeV and a rapidity boost of $\Delta y = 4.8$. The energy of the heavy-ion beams can reach up to 2.76 TeV per nucleon,  yielding $\sqrt{s_{NN}} = 72$~GeV for collisions on a fixed target and a rapidity shift $\Delta y = 4.2$. These conditions necessitate an experimental apparatus that can cover the forward rapidity region and handle a high density of tracks, preferably with the capability for track identification. The LHCb experiment meets all these requirements.

The LHCb detector~\cite{LHCb:2008vvz} is a single-arm spectrometer that provides track reconstruction, charge hadron and muon identification over a wide range of momentum, and a vertex detector allowing for efficient studies of charmed and bottom hadrons using their full reconstruction down to transverse momentum $\pT \approx 0$. It covers the forward rapidity region with geometrical acceptance in pseudorapidity of $2 < \eta < 5$. At LHCb, measurement of $D$ and $B$ mesons allows one to access relatively large values of $x_2$, the momentum fraction of the parton in the target: $0.06 \lesssim x_2 \lesssim 0.76$ for $B$ mesons and $0.02 \lesssim x_2 \lesssim 0.27$ in the case of $D$ mesons.\footnote{The value of $x_2$ is estimated assuming $2 \to 2$ kinematics for the $D$ or $B$ meson production process, which leads to a simple formula: $x_{1,2} =M e^{\pm y_{\rm cms}} / \sqrt{s}$, where $M$ is the mass of the meson. Additional estimates of the probed $x$ region based on NLO calculations are provided in Fig.~\ref{fig:rap-dist}.} Such values of $x_2$ cover the part of EMC and antishadowing regions in the target nucleus, which are difficult to probe effectively in collider experiments.    

LHCb had successfully implemented the first fixed-target program using its System for Measuring Overlap with Gas (SMOG), which was developed for luminosity measurements. In this case, a noble gas was injected into the vertex locator (VELO) system, and the gas served as a target. LHCb collected samples of fixed-target collision data using the LHCb proton and lead beams during the LHC Run 2, performed their analysis and delivered important physics results~\cite{LHCb:2018ygc,LHCb:2018jry,LHCb:2022bbb,LHCb:2022cul,LHCb:2022sxs,LHCb:2022qvj,LHCb:2024vwi,LHCb:2024syd}. The SMOG system provided an excellent demonstration of the feasibility of such studies, but it also had its limitations. The available luminosity was limited by the possible density of gas inside the VELO, and only a few ion species (He, Ne and Ar) were available for experimentation. These shortcomings were addressed with the new system, SMOG2, and the installation of a gas storage cell before the start of LHC Run 3~\cite{BoenteGarcia:2024kba}. The storage cell allows one to inject gas into a well-defined volume within the LHC beam line. It facilitates a higher density of gas within the cell while still maintaining outside the vacuum level required for LHC operations. This upgrade increases the available luminosity in the fixed-target mode by up to two orders of magnitude, allowing for collecting up to 100 pb$^{-1}$ data annually with a proton beam on the H$_2$ target and enabling the use of new ion species, provided that the given gas is compatible with LHC operation. These improvements open new perspectives for measurements of charm- and bottom-quark production in $p$A collisions over a wide range of target atomic mass numbers A with unprecedented precision. In turn, it provides a unique opportunity to significantly improve the current understanding of gluon nPDFs.
%
%--------------------------------
\subsection{Experimental conditions with SMOG2 at LHCb}
\label{subsec:ExpSet}
%--------------------------------
The target gas species and luminosities for SMOG2 that are used in this study were taken from~\cite{Bursche:2018orf} and are summarised in~\ct{tab:Lumin1}. These numbers were proposed as a possible data-taking scenario for Run 3. 
SMOG2 has already performed runs with H$_2$, He, Ne and Ar target. Injections into SMOG2 of gases such as D$_2$, N$_2$, O$_2$, Kr and Xe are being studied.%
\footnote{Recent measurements with a D$_2$ target were reported at the Quark Matter conference~\cite{QM2025}.
}
In our study, that focuses on the atomic mass number dependence of nPDFs, D$_2$ was not considered but it is also of great interest for the nPDFs extraction, in particular to test the isospin symmetry assumption of the nPDFs.

\begin{table}[ht]
\centering
\begin{tabular}{| c | c | c | c |}
\hline
 System & Atomic mass number & $\int\mathcal{L}$ (pb$^{-1}$)\\
 \hline
 $p$H$_{2}$ & 1 & 150 (10$^4$)\\
 $p$He  & 4 & 40\\
 $p$Ne  & 20 & 40\\
 $p$Ar  & 40 & 45\\
 $p$Kr  & 84 & 30\\
 $p$Xe  & 132 & 22 (310)\\
 \hline
\end{tabular}
\caption{Summary of the possible collision systems and luminosities for the SMOG2 experiment for the LHC Run 3~\cite{Bursche:2018orf} at $\sqrtsnn = 113$ GeV. These luminosities are used for $D$ meson pseudo-data generation. The luminosities in parenthesis correspond to those used for $B$ meson pseudo-data generation as proposed in Ref.~\cite{Hadjidakis:2018ifr}.}
\label{tab:Lumin1}
\end{table}
Since the cross sections for $B$ mesons are small, higher luminosities indicated in parenthesis in~\ct{tab:Lumin1} are also considered in this study in the case of H$_2$ and Xe targets, as proposed in Ref.~\cite{Hadjidakis:2018ifr}. 
The rapidity acceptance of the LHCb detectors in the laboratory frame for open heavy-flavor measurements is taken as $2 < y_{\rm lab} < 4.6$~\cite{LHCb:2018jry}, which translates to a center-of-mass (cms) coverage of $-2.79 < y_{\rm cms} < -0.19$, corresponding to $\Delta y \simeq$ 4.79, for the collisions at $\sqrt{s_{NN}} = 113$~GeV with a proton beam energy $E_p=6.8$ TeV.
We consider studies with the $D^0$ meson reconstructed via its hadronic decay $D^0 \to K^- \pi^+$, and $B^{+}$ measured through its decay into $D$ meson: $B^+ \to D^0 + \pi^+$ (and charge conjugate particles $\overline{D}^0$ and $B^-$, respectively). There is a possibility to improve the statistical precision for $B$-meson studies if other decay channels are considered, but combining such measurements with different systematic uncertainties is not always straightforward. Thus, we focus here on a single $B^{\pm}$ decay channel. The overall efficiency for $D$ mesons, that is the product of the geometrical acceptance of the detector and the efficiencies of the reconstruction, the selection, the PID and trigger requirements, is inspired by the SMOG analysis~\cite{LHCb:2022qvj} and from SMOG2 simulations for $D$ mesons for the rapidity dependence~\cite{Santimaria:private}. The integrated efficiency over rapidity is taken as 1\% and it varies as a function of rapidity: from 0.04-0.3\% at the lowest rapidities, reaching a maximum value of 2.1\% at $y_{\rm lab} \approx 3.7$ and decreasing to 0.7\% at $y_{\rm lab} \approx 4.6$. For the $B$ meson, there are not yet studies with the SMOG2 system and we assume a similar overall efficiencies as for the $D$ meson. We also assume the cross sections measurements will have systematic uncertainties of 4-5\% for $D$ mesons and 8\% for $B$ mesons in addition to the statistical ones. Those values are based on previous LHCb measurements in collider mode with large data samples~\cite{LHCb:2017yua, LHCb:2019avm} and the systematic uncertainties expectations on \jpsi cross section measurements with SMOG2 of the order of 2-3\%~\cite{Bursche:2018orf}. In addition, we assume the systematic uncertainties are uncorrelated between the $pp$ and $p{\rm A}$ collision systems.

\section{\label{sec:formalism}Methodology}
\subsection{\label{sec:methodology-overview}Overview of the methodology}

We use the PDF reweighting method to assess the impact of data expected to be collected by the LHCb experiment in a fixed-target mode during LHC Run 3 on the determination precision of gluon nuclear PDF. The study consists of the following steps:
\begin{enumerate}
    \item Computation of the cross section for charm- and bottom-quark production in $p$A reaction at  $\sqrt{s_{NN}}$~=~113~GeV using the \texttt{MadGraph5\_aMC@NLO} framework~\cite{Flore:2025ync}. 
    \item Generation of the pseudo-data for $D^0$- and $B^{+}$- meson production, that reflect the expected statistical and systematic uncertainty anticipated for measurements at LHCb in the fixed-target mode during the LHC Run 3.
    \item Calculation of nuclear modification factors $R_{p{\rm A}}$ with the expected experimental uncertainties.
    \item Application of the PDF reweighting technique to assess what improvement in terms of uncertainty is expected for gluon nPDF if those data are used in the nPDF determination. 
\end{enumerate}

The rest of this section provides details on each of the phases of our study. 

\subsection{\label{sec:pQCD}Cross section for charm- and bottom-quark production}
We calculate the cross section for charm- and bottom-quark production, $\sigma_{p{\rm A} \to c\overline{c}}$ and $\sigma_{p{\rm A} \to b\overline{b}}$, respectively, using the \texttt{MadGraph5\_aMC@NLO} framework~\cite{Alwall:2014hca} (\MG in what follows) extended to accommodate asymmetric reactions~\cite{Flore:2025ync}. \MG allows for automated perturbative QCD calculations for standard model processes at Next-to-Leading Order (NLO) accuracy. The differential cross section $d\sigma_{p{\rm A} \to Q\overline{Q} X}$ (where $Q = c$ or $b$ quark) is given by the general formula:
\begin{equation}
\begin{aligned}
& d\sigma_{p{\rm A} \to Q\overline{Q} X} = &  \\ & A_{\rm A} \sum_{i,j} \int dx_i dx_j \,
f_{i/p}(x_i,\mu_F; {\tt LHAID\_p}) \, f_{j/{\rm A}}(x_j,\mu_F; {\tt LHAID\_A}) \\
&\times
d\widehat{\sigma}_{ij \to Q\overline{Q}}\left(x_i,x_j, \mu_F, \alpha_S(\mu_R; {\tt LHAID\_A})\right),
 \label{eq:factASYM}
 \end{aligned}
\end{equation}
where $A_{\rm A}$ is the atomic mass number of the target nucleus A, $f_{i/p}$ and $f_{j/{\rm A}}$ are the PDFs of the proton and the target nucleus A, $x_i$ and $x_j$ are the fractions of the longitudinal momentum carried by the partons compared to the proton and nucleus A, respectively; $\mu_F$ and $\mu_R$ are the factorisation and renormalisation scales, and $d\widehat{\sigma}_{ij \to Q\overline{Q}}$  is the parton-level cross section for production of heavy quark pair. The {\tt LHAID\_p} and {\tt LHAID\_A} indices indicate that the inputs are taken from the {\tt LHAPDF} library~\cite{Buckley:2014ana}. 

To study the nuclear effects of nPDF on particle production, we use the nuclear modification factor, $R_{p{\rm A}}$:
\begin{equation}
R_{pA} = \frac{1}{A_{A}} \frac{d\sigma_{pA}}{d\sigma_{pp}}, 
\label{eq:RpA}
\end{equation}
$R_{pA}$ is very useful for nPDFs studies since theoretical and experimental uncertainties are expected to cancel to some extent within the ratio. Moreover, the \MG version with asymmetric collisions~\cite{Flore:2025ync} provides both single cross sections for the $pp$ and $p$A collisions and nuclear modification factors themselves. 

In our study, we use a selection of different nPDFs: nCTEQ15HIX (a recent update of the nCTEQ15~\cite{Kovarik:2015cma} fit concentrating on the high-$x$ region)~\cite{Segarra:2020gtj},
nCTEQ15HQ (the newest nCTEQ PDF set that includes many heavy-flavor data from $p$Pb collisions at the LHC)~\cite{Duwentaster:2022kpv},
and EPPS21~\cite{Eskola:2021nhw} (the latest set from EPPS group including data for $D$-meson production from $p$Pb collisions at the LHC). For the proton PDFs, we adopt the corresponding free-proton baseline PDFs~\cite{Owens:2007kp,Hou:2019qau}.

\subsection{\label{sec:pseudo-data}Pseudo-data generation}
The pseudo-data for $D^0$ and $\overline{D}^0$ (simply named $D^{0}$ in the following) production at fixed-target proton-nucleus collisions are obtained using the following equation:
\begin{equation}
\sigma_{pA \to D^{0}} = 2 \sigma_{pA \to c\overline{c}} \, f(c \to D^{0}),
\label{eq:dsigm_D0}
\end{equation}
where $\sigma_{pA} \to c\overline{c}$ is the cross section of the bare charm quark production generated by the \MG code with asymmetric collisions within the SMOG2 experiment rapidity range, and $f(c \to D^0) = 0.542 \pm 0.024$~\cite{Gladilin:2014tba} is the fragmentation factor describing the probability of charm quark to hadronise into $D^0$ meson. The factor $2$ is used to include both $D^0$ and $\overline{D}^0$ production. Note that here and in what follows we take $m_{c} = 1.5$ GeV, and we assume that the rapidity of the single charm quark is equal to the rapidity of the meson~\cite{Cacciari:1996wr}. 
The $D$-meson cross section obtained by \MG describes the SMOG $p$Ne data~\cite{LHCb:2022cul} within the theoretical scale uncertainties, the central values being below the data by a factor of two. 
The expected yield $N_{D^0}$ of observed $D^0$ mesons in the experiment is calculated taking into account the overall measurement efficiency, $\varepsilon$, the branching ratio for the decay channel used in the measurement $Br(D^0 \to K^{-} \pi^{+})= (3.947 \pm 0.09 \pm 0.12)\cdot 10^{-2}$~\cite{ParticleDataGroup:2024cfk}, and the sampled luminosity $\mathcal{L}$:   
\begin{equation}
N_{D^0} = \sigma_{pA \to D^{0}} \, \varepsilon \, Br(D^{0} \to K^{-} \pi^{+}) \, \mathcal{L},
\label{eq:N_D0}
\end{equation}
The luminosities for each collision system are provided in \ct{tab:Lumin1} and are based on Ref.~\cite{Bursche:2018orf}.

Similarly to the $D^0$ case, we generate the $B^+$ pseudo-data by computing the corresponding cross section for $B^+$-meson production in $pA$ collisions as,
\begin{equation}
\sigma_{pA \to B^{+}} = 2  \sigma_{pA \to b\overline{b}} \, f(b \to B^{+}),
\label{eq:dsigm_B}
\end{equation}
where $\sigma_{pA \to b\overline{b}}$ is the cross section of the bare bottom quark production, and the fragmentation factor $ f(b \to B^{+}) = 0.404 \pm 0.006$~\cite{ParticleDataGroup:2024cfk} accounts for the probability of $b$ quark hadronising into a $B^+$ meson. Here and below, we use $m_{b} = 4.75$ GeV. The expected $N_{B^+}$ yield is given by
\begin{equation}
N_{B^+} = \sigma_{pA \to B^{+}} \,\varepsilon \, Br(B^{+} \to \overline{D}^{0} \pi^{+}) \,Br(\overline{D}^{0} \to K^{+} \pi^{-}) \, \mathcal{L},
\label{eq:N_B}
\end{equation}
with $Br(B^{+} \to \overline{D}^{0} \pi^{+}) = (4.61 \pm 0.10)\cdot10^{-3}$~\cite{ParticleDataGroup:2024cfk} and  $Br(\overline{D}^{0} \to K^{+} \pi^{-} )= (3.947 \pm 0.09 \pm 0.12)\cdot 10^{-2}$~\cite{ParticleDataGroup:2024cfk}. The luminosity values are provided in \ct{tab:Lumin1} (values in parenthesis) and are based on luminosity estimations for a cell target similar to SMOG2 and for a full year of data taking with the proton beam~\cite{Hadjidakis:2018ifr}.

\begin{figure*}[ht]
\centering
\subfloat[]{\label{fig:D0-rap-dist}
\includegraphics[width = 7cm, keepaspectratio]{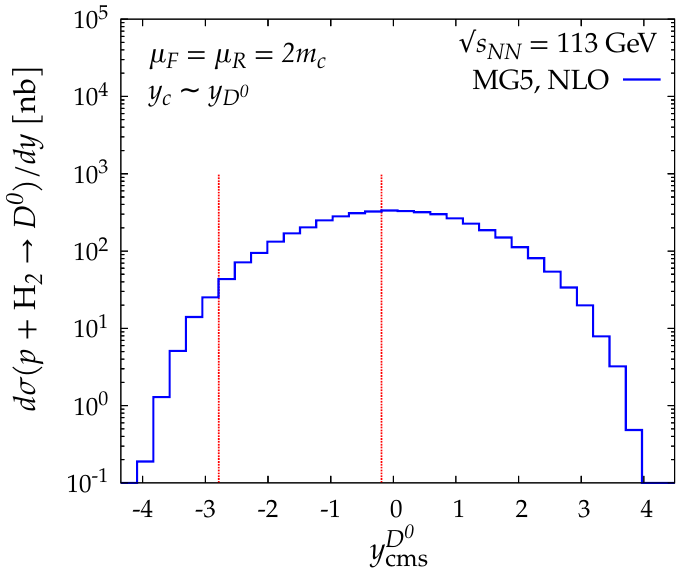}}
\hspace*{5mm}
\subfloat[]{\label{fig:B+-rap-dist}
\includegraphics[width = 7cm, keepaspectratio]{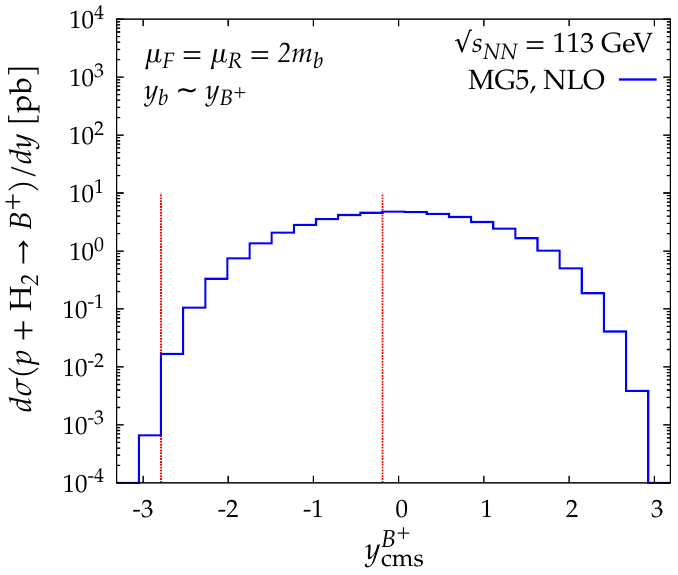}}
\\[-2mm]
\subfloat[]{\label{fig:D0_x2}
\includegraphics[width = 7cm, keepaspectratio]{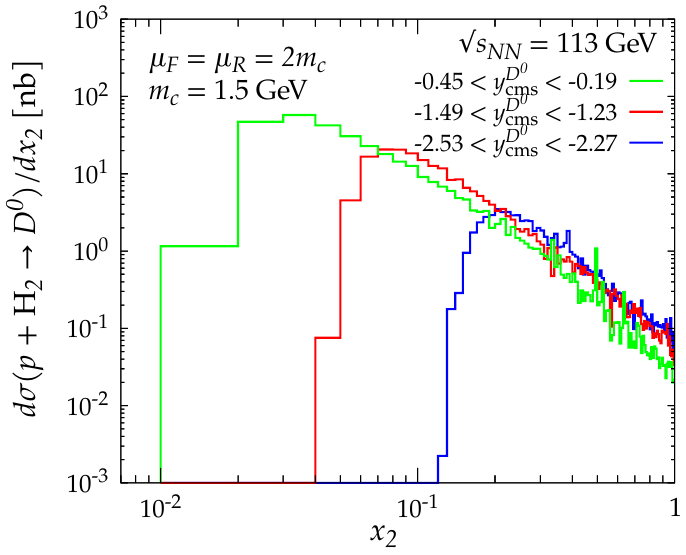}}
\hspace*{5mm}
\subfloat[]{\label{fig:B+_x2}
\includegraphics[width = 7cm, keepaspectratio]{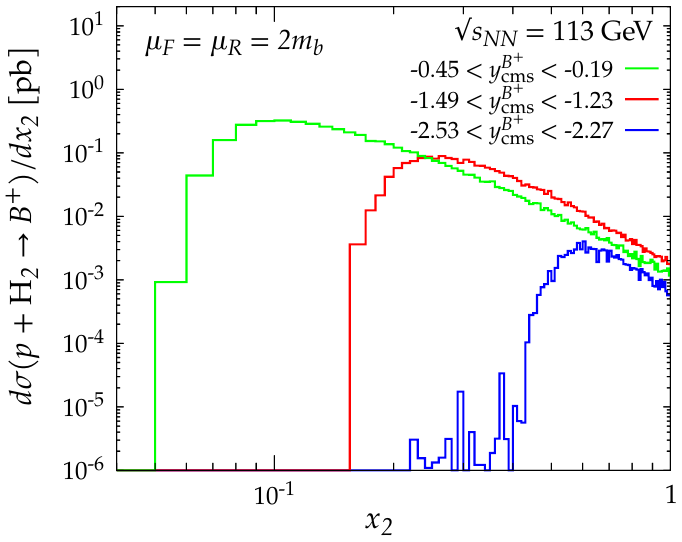}}
\caption{Panels (a) and (b): rapidity distributions of $D^{0}$ and $B^{+}$ mesons  production cross sections in $pH_{2}$ collisions at $\sqrtsnn = 113$~GeV. The red lines show the rapidity coverage expected for SMOG2 at LHCb.  
Panels (c) and (d): $D^{0}$ and $B^{+}$ mesons production cross sections as a function of $x_2$ for three $y_{\rm cms}$ ranges.}
\label{fig:rap-dist} 
\end{figure*}

Figures~\ref{fig:D0-rap-dist} and~\ref{fig:B+-rap-dist} show the cross section for $D^0$ and $B^+$ production in $p$H$_{2}$ reaction at $\sqrtsnn = 113$ GeV as a function of $y_{\rm cms}$. The red lines indicate the LHCb detector acceptance limit in the fixed-target mode. The $B^+$ cross section is about five orders of magnitude below the $D^0$ one and decreases much faster with $|y_{\rm cms}|$ due to its larger mass, which will make $B^+$ measurements at the most backward rapidities ($y_{\rm cms} \lesssim 2.25 $) challenging.
Figures~\ref{fig:D0_x2} and~\ref{fig:B+_x2} present the $D^0$- and $B^+$-mesons production cross sections as a function of $x_2$, the longitudinal momentum fraction carried by partons within the nuclear target, in three selected $y_{\rm cms}$ ranges. Since \MG considers NLO processes, there is a wide distribution of $x_2$ for each rapidity bin. As expected, the $B$-meson production is more sensitive to larger $x$ values as compared to charm mesons. The higher $y_{\rm cms}$ intervals are dominated by lower values of $x$, while the most backward $B^+$ rapidities provide clear access to the large-$x$ region (which is the least known) in the target until $\langle x \rangle \sim 0.68$. Thus, an effort to collect high-precision $B$-meson data within $-2.75 \lesssim y_{\rm cms} \lesssim -2.25$ range within the SMOG2 program will be highly rewarding. 

\subsection{PDF reweighting}
\label{sec:reweighting}

The impact of the SMOG2 pseudo-data on the nPDFs can be studied using the PDF reweighting method~\cite{Giele:1998gw, Ball:2010gb, Sato:2013ika, Paukkunen:2014zia}, which we briefly summarise here. 
The application of this method was initially used in the context of Monte Carlo PDFs~\cite{Ball:2010gb} which are defined via a set of PDF replicas. In such a case central predictions and the corresponding uncertainties for any PDF-dependent quantity, $\cal{O}$, can be computed as average and variance
\begin{equation}
\left\langle {\cal O} \right\rangle = \frac{1}{N_{\rm rep}} \sum^{N_{\rm rep}}_{k=1} {\cal O} (f_{k})\,,
\label{eq:average-unw}
\end{equation}
\begin{equation}
\delta^2\!\left\langle {\cal O} \right\rangle =  \frac{1}{N_{\rm rep}} \sum^{N_{\rm rep}}_{k=1} ({\cal O}(f_{k})- \left\langle {\cal O} \right\rangle)^2 ,
\label{eq:delta-unw}
\end{equation}
where $f_k$ are the PDF replicas and $N_{\rm rep}$ is the total number of PDF replicas.

Through Bayes theorem~\cite{bayes1763lii} the information encoded in the PDFs can be updated with the additional information provided by new data. On a technical level, in the PDF reweighting, this is done by introducing appropriate weights into \ce{eq:average-unw} and \ce{eq:delta-unw}. Taking into account that we will work with Hessian PDF sets, we can define modified Giele-Keller weights as follows~\cite{Paukkunen:2014zia,Kusina:2016fxy}:
\begin{equation}
w_{k} = \frac{e^{-\frac12 \chi^{2}_{k}/T}}{ \frac{1}{N_{\rm rep}} \sum\limits^{N_{\rm rep}}_{i=1} e^{-\frac12 \chi^{2}_{i}/T}},
\label{eq:weights}
\end{equation}
where $T$ is the tolerance adopted in the definition of the Hessian error PDFs, and $\chi^2_k$ is defined as\footnote{In the generation of the pseudo-data, we assume the experimental uncertainties as fully uncorrelated over rapidity so it is acceptable for our analysis to use such a basic definition of the $\chi^2$.}
\begin{equation}
\chi^{2}_{k} =  \sum^{N_{\rm data}}_{i=1} \frac{(D_{i} - T_{i}^{k})^2}{\sigma^{2}_{i}},
\label{eq:chi2}
\end{equation}
where $i$ runs over all the $N_{\rm data}$ pseudo-data points, $D_{i}$ is the $i$-th pseudo-data point, $\sigma_{i}$ is the corresponding pseudo-data uncertainty, and $T_i^k$ is the corresponding theoretical prediction computed with the $k$-th PDF replica.

With the above definitions, we can now compute reweighted PDF-dependent quantities through weighted sum as
\begin{equation}
{\left\langle {\cal O} \right\rangle}_{\rm new} = \frac{1}{N_{\rm rep}} \sum^{N_{\rm rep}}_{k=1} w_{k} {\cal O}(f_{k}),
\label{eq:average-rew}
\end{equation}
\begin{equation}
\delta^2\!{\left\langle {\cal O} \right\rangle}_{\rm new} =  \frac{1}{N_{\rm rep}} \sum^{N_{\rm rep}}_{k=1} w_{k} ({\cal O}(f_{k})- {\left\langle {\cal O} \right\rangle}_{\rm new})^2 .
\label{eq:delta-rew}
\end{equation}

Additionally, to qualitatively assess the efficiency of the reweighting procedure, one can compute the effective number of replicas $N_{\rm eff}$~\cite{Ball:2010gb}:
\begin{equation}
N_{\rm eff} = {\rm exp}\left[ \frac{1}{N_{\rm rep}} \sum^{N_{\rm rep}}_{k=1} w_{k} \log \left(\frac{N_{\rm rep}} {w_k}\right) \right]\,.
\label{eq:Neff}
\end{equation}
$N_{\rm eff}$ provides an estimate of how many replica sets are contributing to the reweighting procedure. If $N_{\rm eff}\ll N_{\rm rep}$, it indicates there are many replicas with small weights providing a negligible contribution to the reweighting procedure. This would signal a potential problem of the procedure, either because the new data is incompatible with the ones used in the original fit or because they provide substantial new information. In both cases, a new global fit is recommended.

In what follows, we will work with Hessian PDFs from nCTEQ~\cite{Segarra:2020gtj,Duwentaster:2022kpv} and EPPS21~\cite{Eskola:2021nhw} fits. We will follow the procedure of Ref.~\cite{Kusina:2016fxy}: before applying the reweighting procedure, we first convert the Hessian PDF sets (represented by central PDF and a set of error PDFs) into PDF replicas, $f_k$, as 
\begin{equation}
f_{k} = f_0 +\sum^{N}_{i=1} \frac{f_i^{(+)} - f_i^{(-)}}{2} R_{ki}
\label{eq:decompALPHA}
\end{equation}
where $f_0$ is the central PDF set, $f_i^{(+)}$ and $f_i^{(-)}$ are the plus and minus error PDFs corresponding to the eigenvector direction $i$, $N$ is the number of Hessian eigenvectors and $R_{ki}$ are random numbers from a Gaussian distribution with mean $0$ and standard deviation of $1$, which are different for each replica $k$ and each eigen-direction $i$.%
\footnote{Such an approach provides a perfect reproduction of PDF errors for Hessian PDF sets using the single tolerance criterion and a symmetric definition of uncertainties~\cite{Kusina:2016fxy}. This is the case for the PDFs from the nCTEQ group. In the case of EPPS21 PDFs, one could use a slightly modified prescription provided in~\cite{Hou:2016sho}. However, since the difference with the asymmetric uncertainty definition is small, we restrict to using the same procedure as in Eq.~\eqref{eq:decompALPHA} (which is equivalent to the Hessian reweighting method employed by the EPPS group~\cite{Paukkunen:2014zia}).}

For our purposes, we generate $N_{\rm rep} = 10^4$. Moreover, we note that the nPDFs we adopt in this work are delivered with uncertainties corresponding to a 90\% confidence level (c.l.). Since our reweighting procedure aims at obtaining new PDFs at a 68\% c.l., the tolerance $T$ in \ce{eq:weights} must be rescaled using the following relation: $T_{68\%} = T_{90\%} / (1.645)^{2}$~\cite{Martin:2009iq}. In the case of nCTEQ, $T_{90\%} = 35$, while in the case of EPPS21, $T_{90\%}=33$. For our purposes, as the two tolerances are very close to each other, we adopt for simplicity $T_{90\%} = 35$, resulting in a rescaled tolerance $T_{68\%} \approx 13$.

We also note that in our analysis the PDF-dependent observable ${\cal O}$ will be given by a nuclear modification factors $R_{pA}$ (as a function of rapidity) given by Eq.~\eqref{eq:RpA}. Since uncertainties on proton PDFs are typically much smaller than the corresponding uncertainties for nuclear PDFs, in what follows we will neglect proton PDF errors.

\section{\label{sec:results}Results}
\renewcommand{\arraystretch}{1.2}
\setlength{\tabcolsep}{0.5em}

\begin{figure*}[hbt!]
\centering
\includegraphics[width=16cm,keepaspectratio]{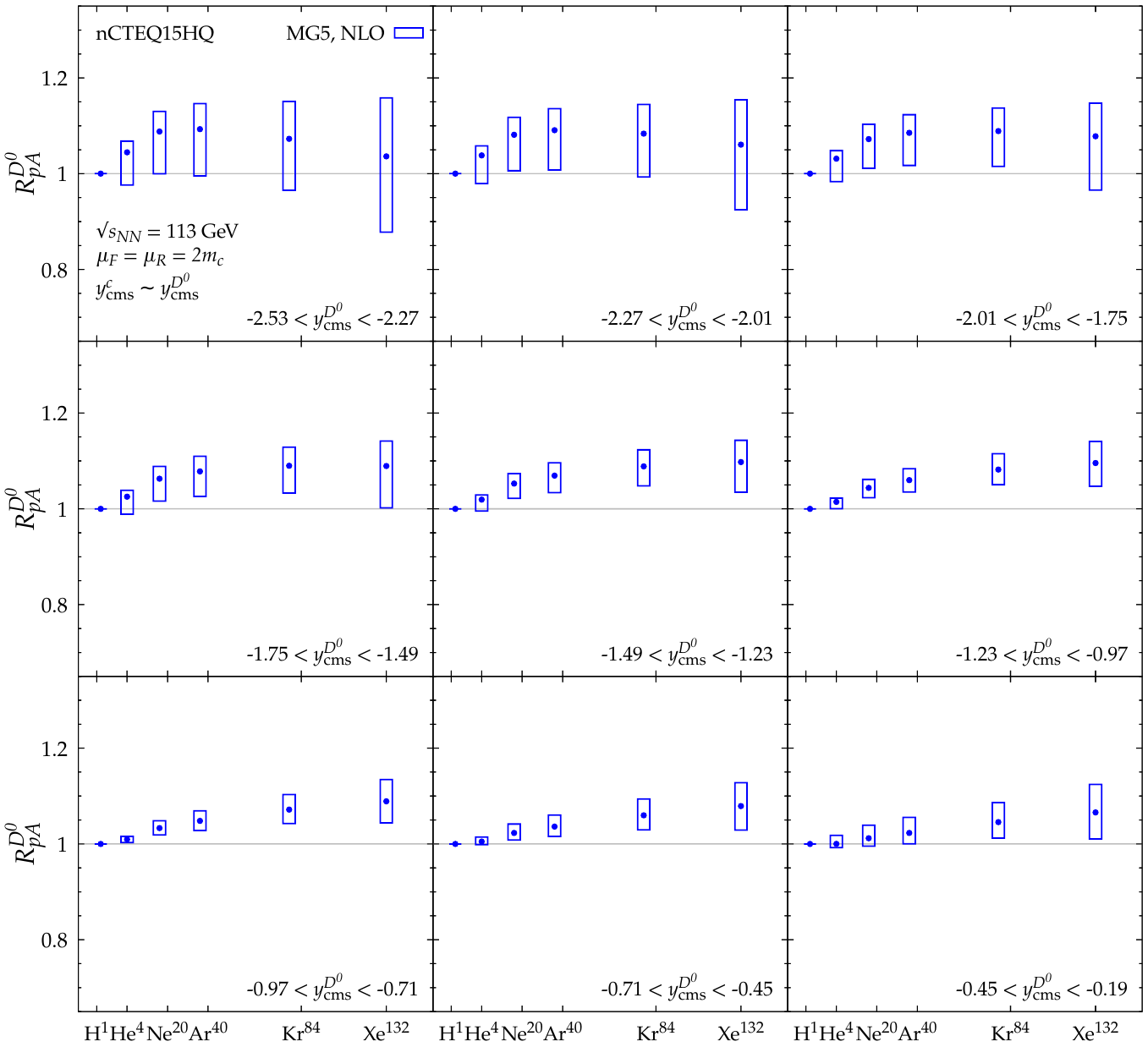}
\caption {
$R_{pA}$ predictions for $D^{0}$ production in proton-nucleus collisions for different nuclei (He, Ne, Ar, Kr, Xe) as a function of the atomic mass number for each considered $y_{\rm cms}^{D^0}$ interval within the LHCb detector acceptance. The nPDF uncertainties from nCTEQ15HQ~\cite{Duwentaster:2022kpv} are shown with blue boxes.
}
\label{fig:A-scan} 
\end{figure*}

We begin by demonstrating the current nPDF uncertainties for $D^0$-meson production process. In the kinematic region of interest, the scale uncertainty for $R_{pA}$ nearly cancels between numerator and denominator, and is negligible compared to the PDF errors. Hence, in what follows, we will only concentrate on the nPDF uncertainty and we will not vary the scale. Figure~\ref{fig:A-scan} shows the nuclear modification factors $R_{pA}$ for He, Ne, Ar, Kr and Xe nuclei for various $D^{0}$-meson rapidity intervals. The boxes represent the current nPDF uncertainties (we adopt here the nCTEQ15HQ nPDFs~\cite{,Duwentaster:2022kpv} as an example). They are the largest for the lowest values of rapidity, which corresponds to relatively large $x$, which corresponds to the end of the EMC region and the beginning of the antishadowing region. Clearly, the uncertainties due to nPDFs are far from being satisfactory. It can be also seen from Fig.~\ref{fig:A-scan} that the uncertainties are larger for heavier nuclei: this is partially driven by the nPDF parameterisations which typically underestimate the uncertainties for light nuclei.

We will use the pseudo-data generated according to the procedure presented in Section~\ref{sec:pseudo-data} and the PDF reweighting method described in Section~\ref{sec:reweighting} to see the potential constraints that can be provided by the future LHCb SMOG2 measurement of $D^0$ and $B$ meson on the current nPDFs. For this purpose, we use a selection of recent nPDFs: EPPS21~\cite{Eskola:2021nhw}, nCTEQ15HQ~\cite{Duwentaster:2022kpv}, and nCTEQ15HIX~\cite{Segarra:2020gtj}. However, before presenting the numerical results, we would like to point out some caveats of such an analysis, which prevents us from showing the full potential of the fixed-target LHCb data. First of all, as already said earlier the uncertainties of gluon nPDFs at large-$x$ are underestimated. This is caused by the combination of two issues: (i) a small amount of available data that allows to constrain this region, and (ii) the limits of the Hessian method, used to obtain nPDF uncertainties, to provide realistic error estimates in the absence of data constraints. Furthermore, in most cases the uncertainties for light-nuclei are systematically underestimated due to the way the nPDF parameterisations are constructed.
Finally, what is probably the most important limitation is that in order to produce pseudo-data for this study we need to base them on theoretical predictions using nPDFs. Typically, this is not a problem, as the crucial part is to have realistic estimates of the uncertainties of pseudo-data. However, in the current study this also means that the pseudo-data we will use encodes the nuclear $A$-dependence of the corresponding nPDFs. Hence, our analysis has limited capability to study the $A$-dependence of nPDFs. Nevertheless, we would like to highlight that the knowledge of the $A$-dependence is very limited, therefore the future SMOG2 measurements will give a unique opportunity to determine it and to test the assumptions that are currently used in nPDF fits to model it.

In what follows, we will consider two scenarios for the PDF reweighting: 
(i) {\em scenario 1}: where we perform the reweighting of nPDFs on given nucleus using only the pseudo-data for this nucleus, and
(ii) {\em scenario 2}: where we use pseudo-data for all considered nuclei (He, Ne, Ar, Kr, Xe) together.
In both scenarios we use the pseudo-data for $D$ and $B$ mesons separately.
On the one hand, {\em scenario 1} is very conservative, as in reality when performing the nPDF fit one would use all the data together to exploit the full constraining power of the experimental measurements. On the other hand, in the reweighting approach we do not have any means to model the $A$-dependence of nPDFs, and in actual fits the constraints from measurements on different nuclei will enter through the $A$-dependence of nPDFs and will test and constrain it, which should be closer to {\em scenario 2}.
\begin{figure*}[t]
\centering
\includegraphics[width=16.5cm, keepaspectratio]{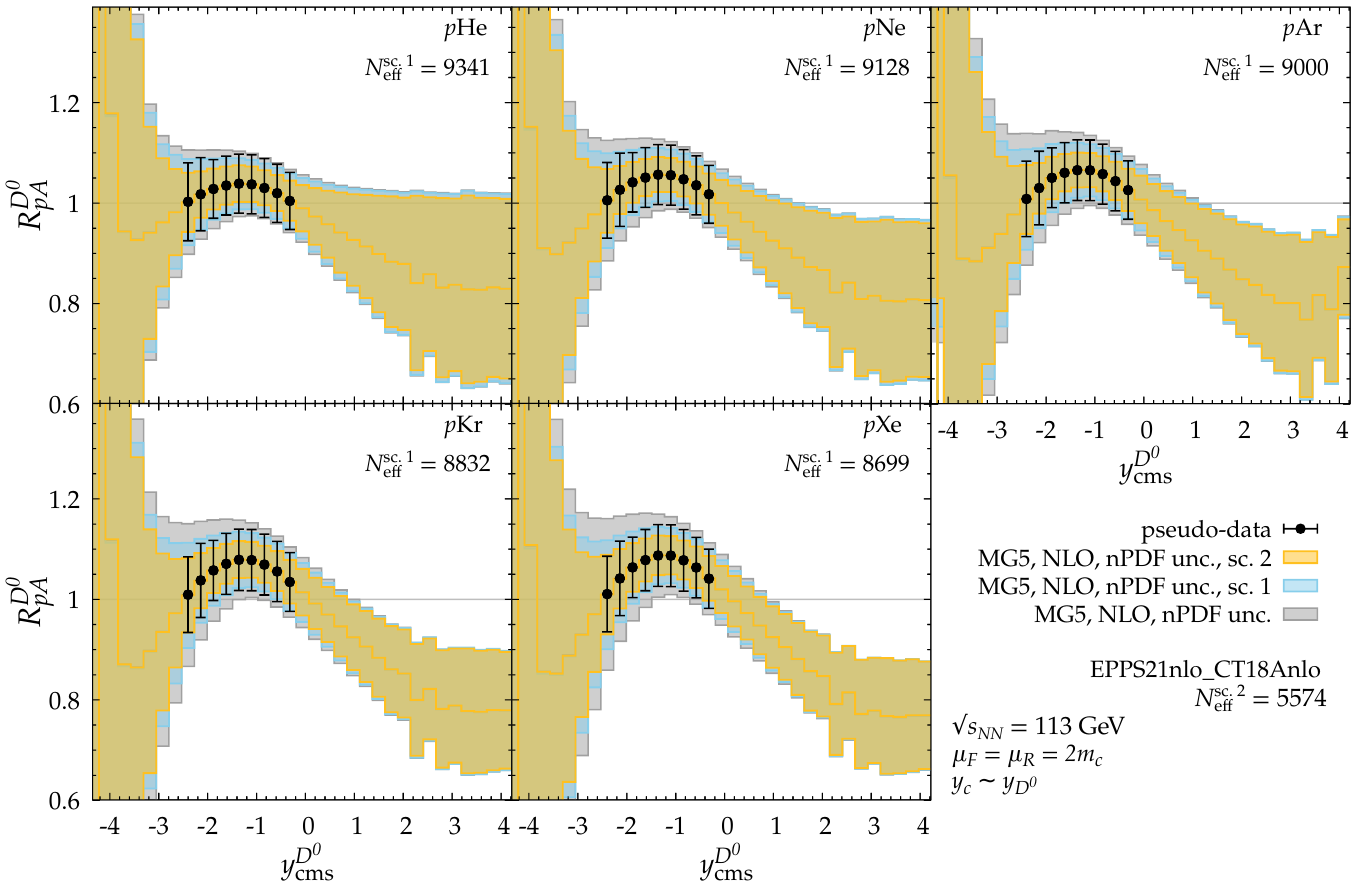}%}
\caption {$R_{pA}$ predictions for the $D^{0}$-meson production in $p$A collisions for different nuclei (He, Ne, Ar, Kr, Xe) for the EPPS21 nPDFs~\cite{Eskola:2021nhw}. Pseudo-data are shown in black, while the nPDF uncertainties before and after the reweighting for scenarios 1 and 2 are shown with grey, blue and yellow bands respectively.
}
\label{fig:D0_all_nucl} 
\end{figure*}
\begin{figure*}[htbp]
\centering
\includegraphics[width=16.5cm,keepaspectratio]{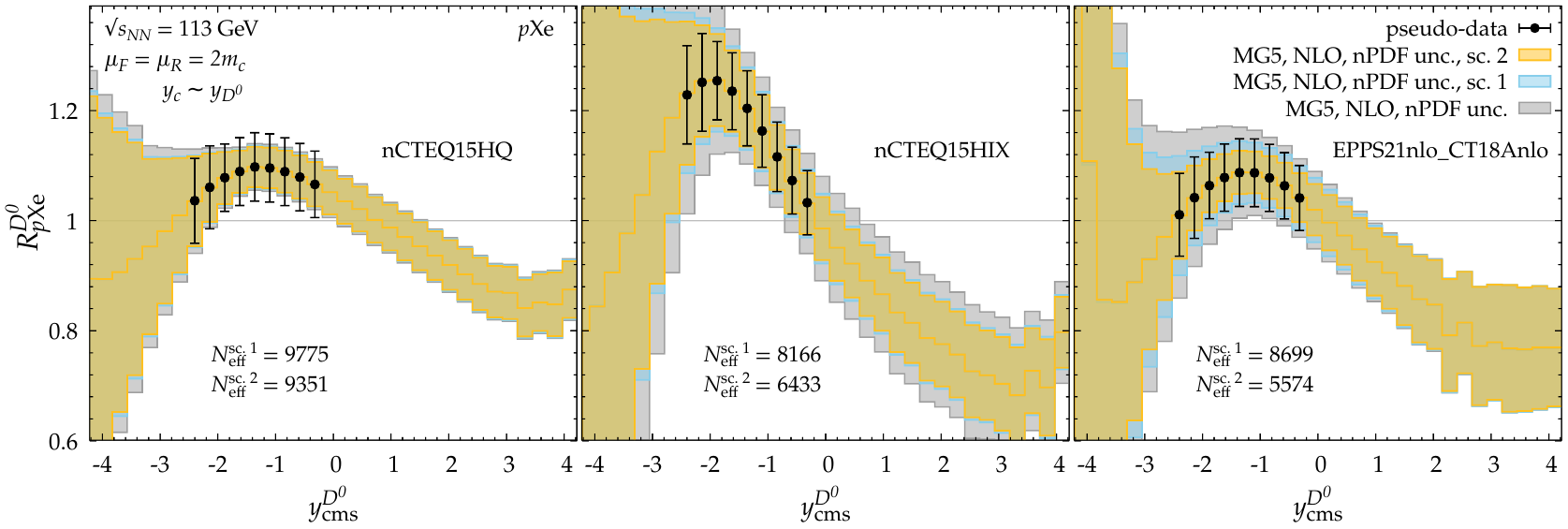}
\caption {$R_{p{\rm Xe}}$ predictions for the $D^{0}$-meson production in $p$Xe collisions for different sets of nPDFs (nCTEQ15HQ~\cite{Duwentaster:2022kpv}, nCTEQ15HIX~\cite{Segarra:2020gtj}, EPPS21~\cite{Eskola:2021nhw}). Pseudo-data are shown in black, while the nPDF uncertainties before and after the reweighting are shown with grey, blue and yellow bands respectively.
}
\label{fig:D0_all_PDFs} 
\end{figure*}

Figure~\ref{fig:D0_all_nucl} shows $R_{pA}$ as a function of the rapidity of the $D^{0}$ meson for the corresponding nuclei, both before and after the reweighting using the two scenarios. We adopt a conservative approach and assume  uncorrelated systematic uncertainties between the $p$H and $p$A data samples to compute the $R_{pA}$ pseudo-data uncertainties. We use here the EPPS21 nPDFs as a representative example, while later we provide additional comparisons with results obtained with different nPDFs. The SMOG2 pseudo-data cover midrapidity to moderate backward rapidity region in the cms frame, which in the fixed-target mode is the region where the antishadowing effect is expected (the corresponding $x$ values can be read from Fig.~\ref{fig:D0_x2}). This fact is reflected in Fig.~\ref{fig:A-scan} where the values of $R_{pA}$ for all rapidity intervals are higher than unity.

The pseudo-data in Fig.~\ref{fig:D0_all_nucl} are shown with black error bars, while the nPDF uncertainties before and after the reweighting using {\em scenario 1} and {\em scenario 2} are represented by grey, blue and yellow bands, respectively. The number of effective replicas, $N_{\rm eff}$, in the two scenarios is also indicated in the plots. From the figure, we can conclude that the reweighting method works well with the selected SMOG2 $D^{0}$-meson pseudo-data. Even though these data cover only a relatively narrow range in rapidity, they will have a significant impact on nPDFs uncertainties also outside of the probed rapidity interval. The precision of nPDF determination is improved for the EMC ($-4\lesssim y_{\rm cms} \lesssim -2$) and even shadowing ($0.5\lesssim y_{\rm cms} \lesssim 4$ ) regions. 

We further concentrate on the $p$Xe distributions, for which we expect to have the most realistic nPDF uncertainty estimates. In Fig.~\ref{fig:D0_all_PDFs} we display $R_{p\mathrm{Xe}}$ for three nPDF sets: nCTEQ15HQ (left), nCTEQ15HIX (middle) and EPPS21 (right) for both scenarios. The reason for choosing these nPDFs is that the nCTEQ15HQ nPDFs includes most constraints on the nuclear gluon distribution because of the inclusion of a broad range of $D$- and $B$-meson and quarkonium data from $p$Pb collisions at the LHC. These are predominantly sensitive to the low-$x$ gluon PDF but through sum rules and the parameterisations also impact higher-$x$ values. On the other hand, the nCTEQ15HIX nPDFs provide a dedicated treatment of the high-$x$ region -- in particular the EMC region -- which is of interest for the current study and should allow for a more realistic estimate of nPDF uncertainties in this region (especially for quarks).
When comparing with the results for EPPS21 nPDFs we can see that the largest reduction of the uncertainties occurs in the same region of negative rapidities (high $x$). However, for the nCTEQ15HIX the reduction also extends to the positive rapidities which is related to the absence of the constraints for low-$x$ gluons coming from the collider heavy-quark(onium) data. This is not the case for the nCTEQ15HQ, for which these kind of data were included in the fit.

\begin{figure*}[t]
\centering
\includegraphics[width=16.5cm,keepaspectratio]{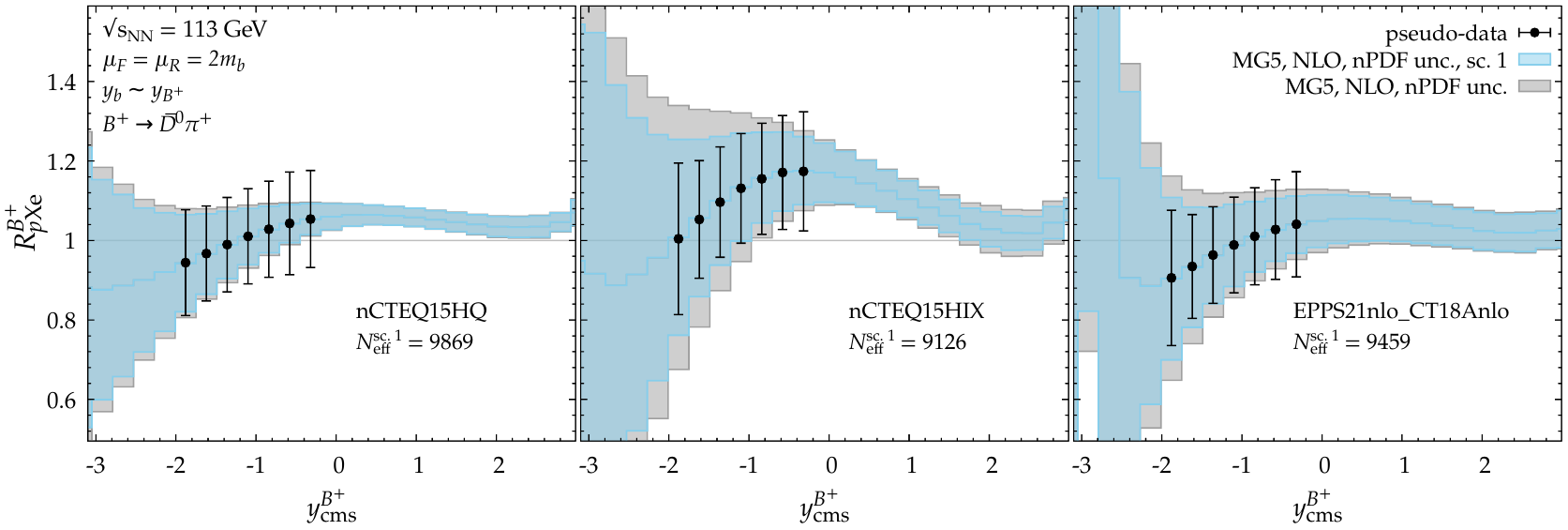}
\caption{$R_{p{\rm Xe}}$ predictions for $B^{+}$-meson production in $p$Xe collisions for different sets of nPDFs (nCTEQ15HQ~\cite{Duwentaster:2022kpv}, nCTEQ15HIX~\cite{Segarra:2020gtj}, EPPS21~\cite{Eskola:2021nhw}). Pseudo-data are shown in black, while the nPDF uncertainties before and after the reweighting are shown with grey and blue bands respectively.}
\label{fig:B-all-PDFs} 
\end{figure*}

In Fig.~\ref{fig:B-all-PDFs} we present the results of the reweighting using $R_{pA}$ as a function of the rapidity of the $B^{+}$ meson in $p$Xe collisions for the same set of nPDFs. Pseudo-data are shown in black, and nPDF uncertainties before and after the reweighting are shown as grey and blue bands, respectively. 
In this case, we restrict the reweighting to the pseudo-data for Xe only (corresponding to {\em scenario 1} for the $D$ meson). Notice that the assumed luminosities for $B$ mesons are larger than the ones for $D$ mesons (see Tab.~\ref{tab:Lumin1} and Section~\ref{subsec:ExpSet}).\footnote{For the $B$-meson measurements, uncertainties will be limited by the statistics, whereas for the $D$ meson they are expected to be limited by the systematic uncertainties.}
The results exhibit similar features as for the $D$ mesons, as we observe a reduction of uncertainties in the large negative rapidity region. However, in the case of $B^+$ mesons, the probed $x$ values are larger, getting closer to the EMC region (we indeed start to see suppression in $R_{p\mathrm{Xe}}$). This is something that is reflected in the results for the nCTEQ15HIX nPDFs. Here the uncertainties at larger negative rapidities are reduced more than in the case of $D$ meson (see Fig.~\ref{fig:D0_all_PDFs}), and the reduction is more prominent when the rapidity is more negative. Even though the probed rapidity range and the expected measurement accuracy are limited, the $B^+$ data will allow a significant improvement in our description of the EMC effect for gluons. 

\begin{figure*}[htbp]
\centering
\includegraphics[width=16.5cm,keepaspectratio]{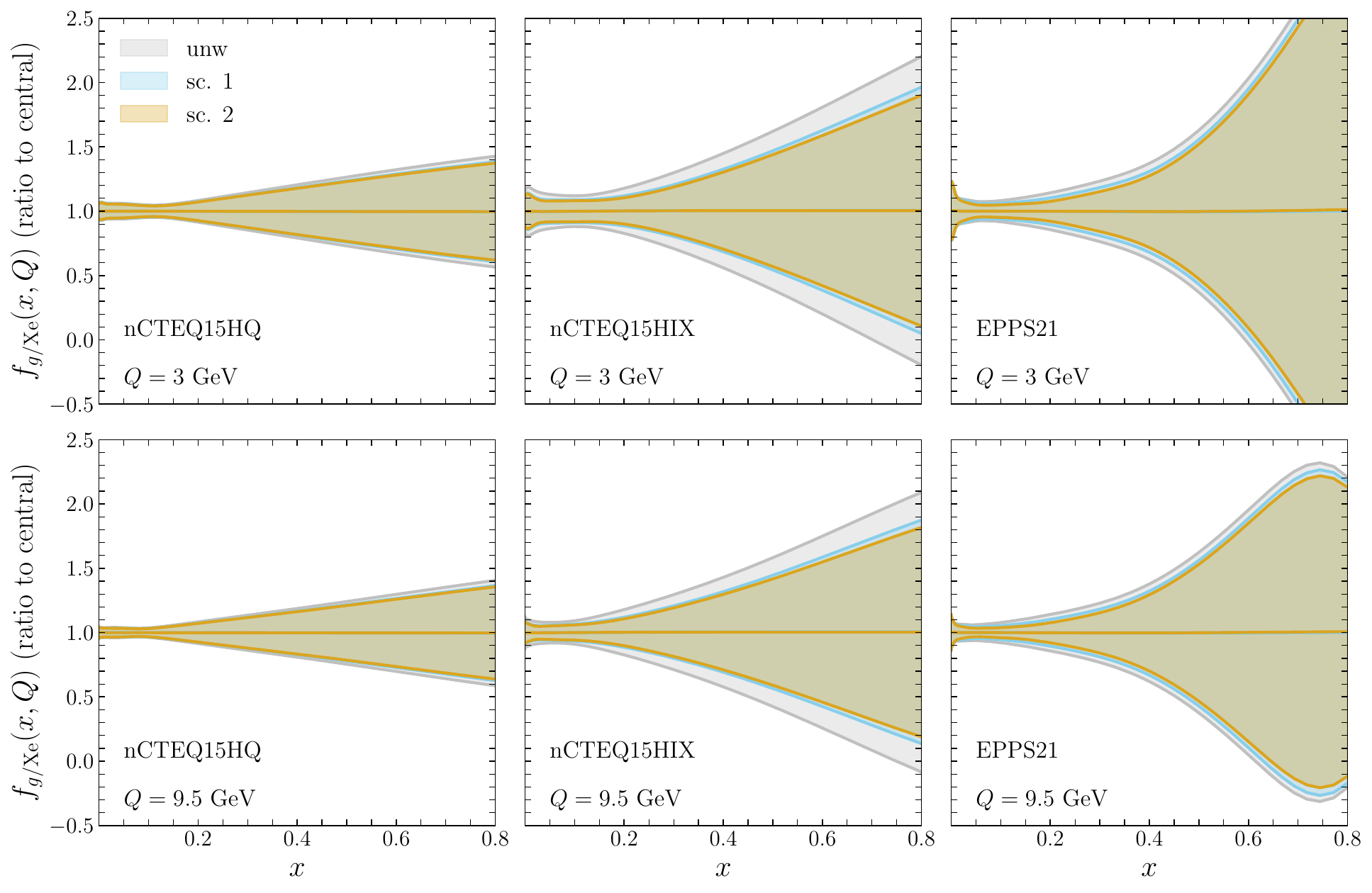}
\caption{Gluon nPDF for Xe nucleus for nCTEQ15HQ~\cite{Duwentaster:2022kpv} and EPPS21~\cite{Eskola:2021nhw} at $Q = 3$ GeV and $Q = 9.5$ GeV.}\label{fig:gluon-Xe-nPDF-linscale} 
\end{figure*}
\begin{figure*}[htbp]
\centering
\includegraphics[width=16.5cm, keepaspectratio]{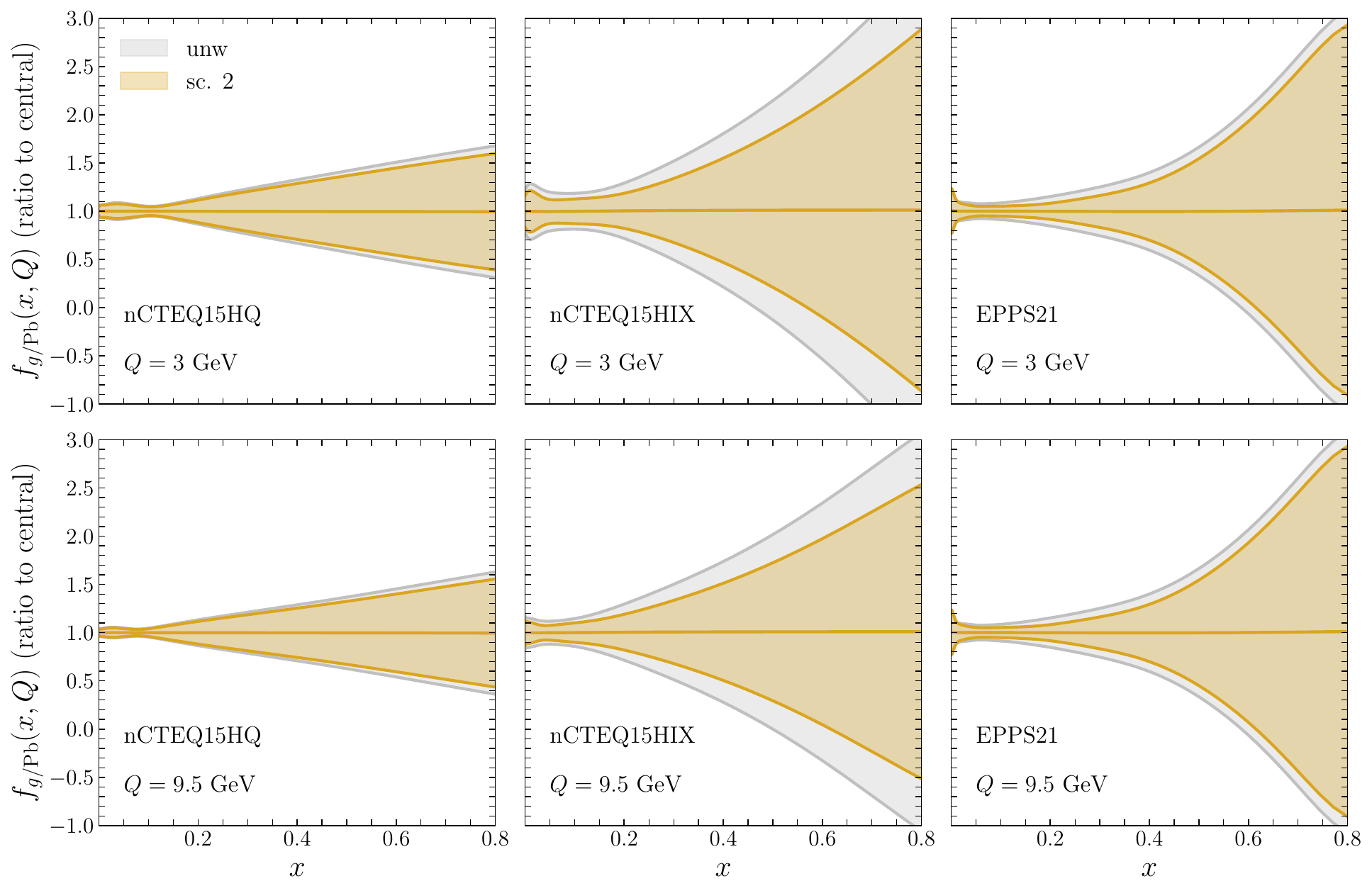}
\caption{Gluon nPDF for Pb nucleus for nCTEQ15HQ~\cite{Duwentaster:2022kpv} and EPPS21~\cite{Eskola:2021nhw}  at $Q = 3$ GeV and $Q = 9.5$ GeV.}\label{fig:gluon-Pb-nPDF-linscale} 
\end{figure*}
\begin{figure*}[t!]
\centering
\includegraphics[width=16.5cm,keepaspectratio]{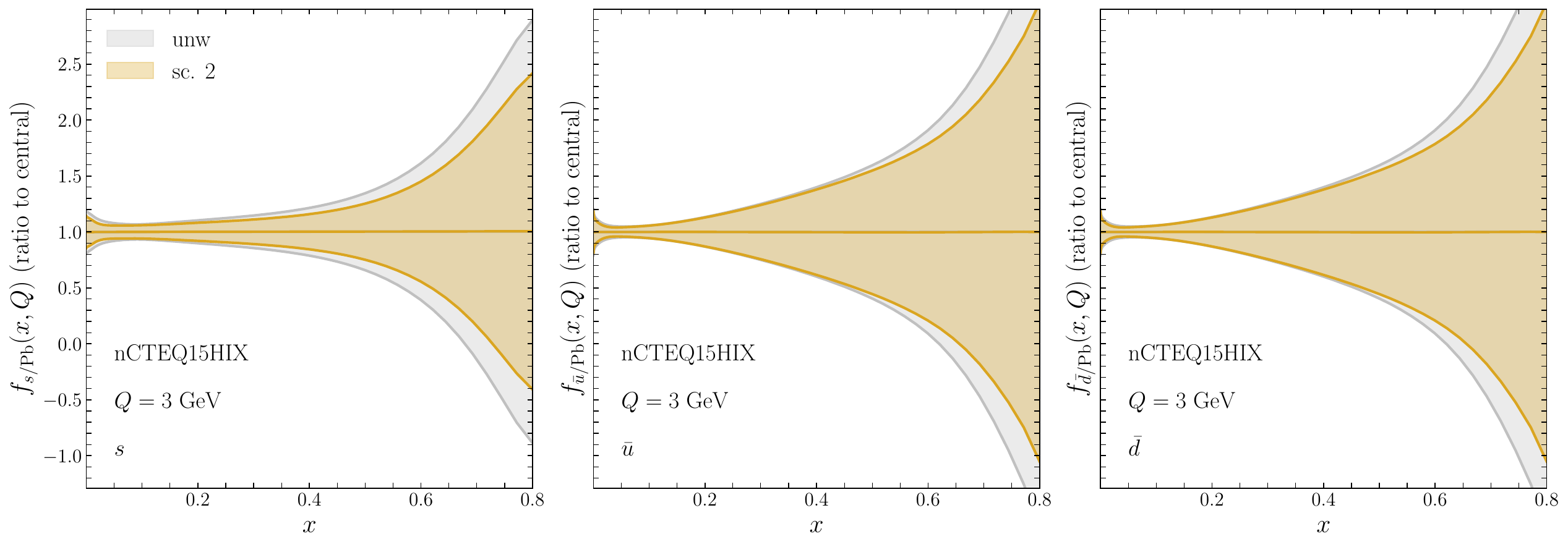}
\caption{Strange-, anti-up-, and anti-down-quark distributions in Pb nucleus from the reweighting with nCTEQ15HIX nPDFs~\cite{Segarra:2020gtj} at $Q = 3$ GeV.}\label{fig:quark-Pb-nPDF-linscale} 
\end{figure*}
As a last step, we investigate directly the reduction of uncertainties at the level of nPDFs. We concentrate on the gluon nPDF, which dominates the production mechanism of $D$ and $B^+$ meson, and as such is the flavour that is the most constrained by these data.
In Fig.~\ref{fig:gluon-Xe-nPDF-linscale} we present the gluon nPDFs for Xe nucleus obtained by the reweighting using the $D^0$ meson data in scenarios 1 and 2. We show here normalised ratios for nCTEQ15HQ, nCTEQ15HIX and EPPS21 nPDFs for two values of factorisation scale,  $\mu_F = 3$ GeV and $\mu_F = 9.5$ GeV. The plots are done using a linear scale to highlight the high-$x$ region of the distributions. We can see that indeed for all PDFs the uncertainties in this region are reduced. In the case of nCTEQ15HQ the reduction occurs for $x\gtrsim0.35$ and is rather small, for EPPS21 and nCTEQ15HIX it starts already for $x\gtrsim0.1$ and is larger in magnitude, especially in the case of nCTEQ15HIX. The uncertainties in the low-$x$ region are not affected, except for the nCTEQ15HIX case.

To give further perspective to the observed reduction of nPDF uncertainties, we will also consider the impact of the pseudo-data on the nPDFs of lead nucleus. The LHCb experiment in the fixed-target mode with SMOG2 will not be able to take data on lead. However, in analogy to what we did in scenario 2, when using pseudo-data for different nuclei to assess the impact on Xe nPDF, we can apply Eqs.~\eqref{eq:average-rew} and~\eqref{eq:delta-rew} and use the same data to see the potential impact on the Pb nucleus.
This is particularly interesting, as with the advent of $p$Pb collider data from the LHC, the nPDFs of lead are actually the best constrained distributions with the most reliable estimates of their uncertainties.
Therefore, in Fig.~\ref{fig:gluon-Pb-nPDF-linscale} we show the reduction of gluon uncertainties in Pb nucleus for nCTEQ15HQ, nCTEQ15HIX and EPPS21 nPDFs after reweighting with pseudo-data for $D^0$ meson in scenario 2. The results are shown for two values of factorization scale: $\mu_F = 3$ GeV and $\mu_F = 9.5$ GeV.
Similarly to Fig.~\ref{fig:gluon-Xe-nPDF-linscale}, we observe a reduction of high-$x$ uncertainties for all the considered nPDFs and starting at similar values of $x$. However, the reduction in the case of lead gluon PDF is larger than for the Xe gluon distribution.
This highlights the point that even with the relatively broad range of collider $p$Pb LHC data providing comprehensive constraints for the gluon Pb nPDF\footnote{The $p$Pb LHC data currently used in nPDF fits and providing constraints on gluon distribution include: $D$- and $B$-meson data, quarkonium data, single inclusive hadron production data, isolated photon data, Drell-Yan and $Z/W$ production data.}
the fixed-target experiment with the LHC beam will allow for a considerable reduction of the current nPDF uncertainties. 
Finally, we check if the considered $D$- and $B$-meson data can have any impact on the quark nPDFs. For this purpose we look at the results obtained with the nCTEQ15HIX nPDFs, as the dedicated treatment of the high-$x$ region in that analysis allows for a more realistic estimate of the high-$x$ behavior, especially in the case of quark distributions. In Fig.~\ref{fig:quark-Pb-nPDF-linscale} we show the strange-, anti-up-, and anti-down-quark distributions in lead nucleus at $\mu_F = 3$ GeV. Even though the considered data is predominantly sensitive to the gluon distribution, we observe that for the strange nPDF the obtained reduction of uncertainties is substantial for $x\gtrsim0.3$. In the case of the $\bar{u}$- and $\bar{d}$-quark distributions the reduction is also present but more moderate and it starts at larger $x$ values ($x\gtrsim0.5$). This reflects the fact that the constraints for the strange nPDF are currently very limited. We also checked that similar results are obtained for the light sea-quarks nPDFs in Xe nucleus.

\section{\label{sec:conclusions}Conclusions}
In this work we adopted the PDF reweighting method to estimate the potential impact of $D$- and $B$-meson production measurements at the fixed-target program, SMOG2, at LHCb experiment on nuclear PDFs (nPDFs). We have shown that such a program provides an excellent opportunity to improve our understanding of gluon nPDF and that it can even provide some constraints for light-quarks nPDFs. The data expected to be collected at SMOG2 have the potential to significantly enhance the precision of gluon nPDF determination, especially at relatively large longitudinal momentum fraction $x$, where the antishadowing and EMC effects are expected ($0.02\lesssim x\lesssim0.8$). This region is poorly known, which is reflected in significant uncertainties of nPDFs. At the same time, it is crucial for constraining hadron structure models, learning the properties of QCD confinement, or explaining the EMC effect~\cite{Higinbotham:2013hta,Malace:2014uea}.

Furthermore, the fixed-target data for a wide range of nuclear targets will provide a unique opportunity to perform the first systematic study of the atomic mass dependence of nuclear PDFs. Such a dependence is indeed currently \textit{assumed} by the nPDFs groups, and SMOG2 measurements will allow for the first earnest test of these assumptions.
Particularly important will be the opportunity to study gluon distribution in light nuclei, such as Ne, which are currently unconstrained.
Moreover, measurements on H$_2$ target will provide a baseline for constructing nuclear modification ratios, $R_{pA}$, allowing for the cancellation of different theoretical uncertainties.
These advancements will, in turn, result in significantly improved precision of pQCD calculations in different areas of particle and nuclear physics, and astrophysics.

\section*{Acknowledgements}
We thank Frederic Fleuret, Emilie Maurice, and Marco Santimaria for useful discussions. We also thank Jean-Philippe Lansberg for his contributions during the early stages of this work.

This project has received funding from the French CNRS via the COPIN-IN2P3 bilateral agreement. 
C.F.~is supported by the European Union ``Next Generation EU'' program through the Italian PRIN 2022 grant n.~20225ZHA7W.
C.H.~acknowledges support by P2IO, Paris-Saclay University. 
D.K.~was supported in part by the National Science Centre, Poland, under the research project no.~2018/30/E/ST2/00089.
A.K.~acknowledges the support of the National Science Centre Poland under the Sonata Bis Grant No.~2019/34/E/ST2/00186.
The work of A.S.~was partially supported by the Excellence Initiative: Research University at Warsaw University of Technology.

\bibliographystyle{elsarticle-num} 
\bibliography{references}

\end{document}